\documentclass[aps,prl,amsmath,amssymb,floatfix,twocolumn, amsmath, superscriptaddress, twocolumn]{revtex4}
\usepackage{multirow}
\usepackage{bbold}
\usepackage{subfigure}
\usepackage{color}
\usepackage{mathrsfs}
\usepackage{hyperref}

\usepackage{amsfonts, relsize, color}
\usepackage{graphics}
\usepackage{graphicx}
\usepackage{subfigure}
\usepackage{hyperref}
\usepackage{color}
\usepackage{gensymb}

\begin{document}
\title{Unconventional superconductivity in nearly flat bands in twisted bilayer graphene
}

\author{Bitan Roy}
\affiliation{Max-Planck-Institut f\"{u}r Physik komplexer Systeme, N\"{o}thnitzer Stra. 38, 01187 Dresden, Germany}

\author{Vladimir Juri\v ci\' c}
\affiliation{Nordita, KTH Royal Institute of Technology and Stockholm University, Roslagstullsbacken 23,  10691 Stockholm,  Sweden}

\date{\today}

\begin{abstract}
Flat electronic bands can accommodate a plethora of interaction driven quantum phases, since kinetic energy is quenched therein and electronic interactions therefore prevail. Twisted bilayer graphene, near so-called the ``magic angles", features \emph{slow} Dirac fermions close to the charge-neutrality point that persist up to high-energies. Starting from a continuum model of slow, but strongly interacting Dirac fermions, we show that with increasing chemical doping away from the charge-neutrality point, a time-reversal symmetry breaking, valley pseudo-spin-triplet, topological $p+ip$ superconductor gradually sets in, when the system resides at the brink of an anti-ferromagnetic ordering (due to Hubbard repulsion), in qualitative agreement with recent experimental findings. The $p+ip$ paired state exhibits quantized spin and thermal Hall conductivities, polar Kerr and Faraday rotations. Our conclusions should also be applicable for other correlated two-dimensional Dirac materials.  
\end{abstract}

\maketitle

\emph{Introduction:} Carbon, owing to its flexibility in chemical bonding, yields a variety of low-dimensional allotropes: fullerene, nanotubes, one-atom thin honeycomb crystal- graphene~\cite{CNT-book, katsnelson-book}. Due to the van der Waals interaction, next generation of allotropes can be synthesized by stacking a few carbon layers~\cite{Grigorieva-Geim-Nature2013}. For example, monolayer graphene features pseudo-relativistic Dirac fermions at low-energies~\cite{graphene-discovery}, responsible for its unusual electronic properties~\cite{ballistic-graphene,Kim-qhe-monolayer}. Furthermore, Bernal arrangement of two graphene layers hosts quadratically dispersing gapless chiral excitations~\cite{bilayer-qhe}, while in rhombohedral trilayer graphene quasiparticles display cubic dispersion~\cite{zhang2011}, falling outside the realm of standard Fermi liquid.

\begin{figure}[t!]
\includegraphics[width=0.95\linewidth]{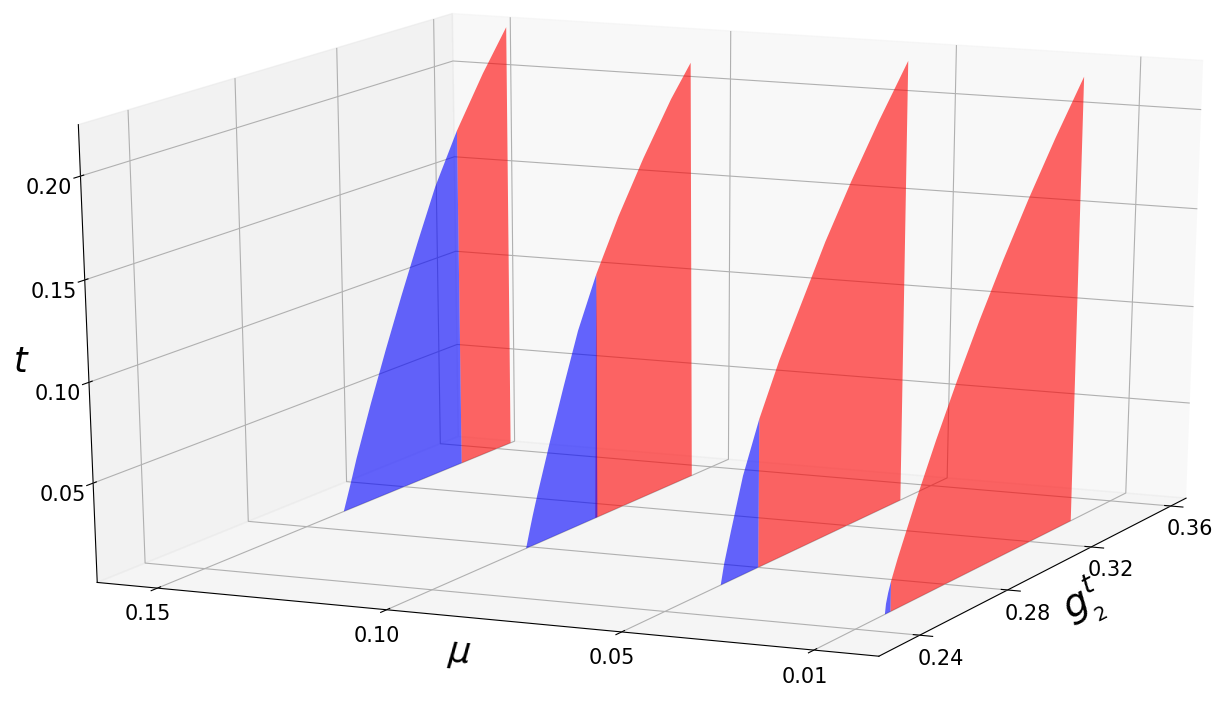}
\caption{Various cuts of the phase diagram of interacting Dirac fermions at finite temperature ($t$) and chemical potential ($\mu$, measured from the CNP), measured in units of $v_F \Lambda$. Here, $g^t_2= U \Lambda/(12 v_F)$ is the dimensionless coupling constant, $v_F$ is the Fermi velocity, $\Lambda$ is the ultraviolet momentum cut-off for slow Dirac fermions, and $U$ is the strength of repulsive onsite Hubbard interaction. The red shaded region is occupied by anti-ferromagnet and the blue region denotes $E_g$ or $p+ip$ pairing, see also Fig.~\ref{phasediagram}. The region outside these shaded ones is occupied by a correlated Dirac liquid without any long-range order. The results are obtained from a renormalization group calculation summarized in Eq.~(\ref{RG:finiteT}) and the Supplementary Materials~\cite{supplementary}. 
}~\label{RG:phaseDiagram}
\end{figure}

\begin{figure}[t!]
\includegraphics[width=4.2cm,height=3.75cm]{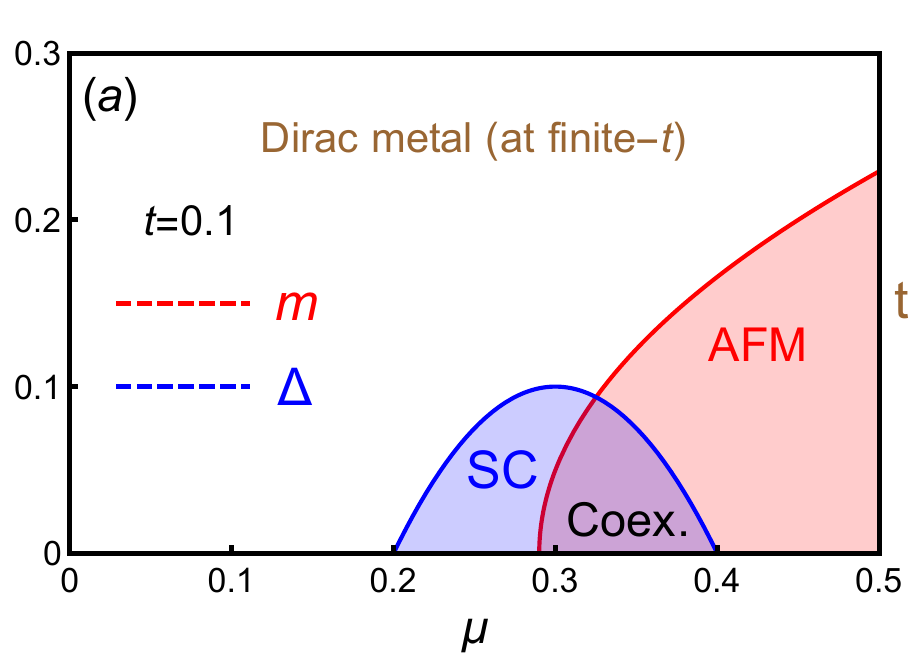}
\includegraphics[width=4.2cm,height=3.60cm]{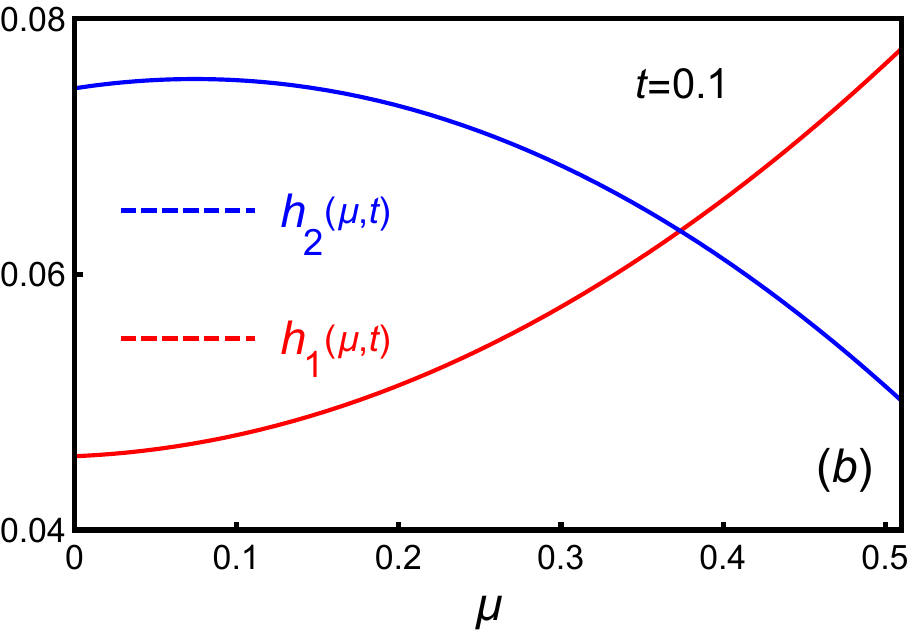}%
\caption{ (a) Amplitudes of anti-ferromagnet ($m$) and $E_g$ pairing ($\Delta$), obtained by minimizing the free-energy [see Eq.~(\ref{free-energy:Eq})], as a function of small to moderate chemical doping ($\mu$), measured from the CNP, at a fixed temperature $t=0.1$. All quantities are dimensionless (see text). Corresponding variation of the coupling constants, $g_{_j}=(U \Lambda/v_{_F})\; h_j (\mu,t)$ in these two channels, for $j=1,2$, where $h_j (\mu,t)$ are phenomenological functions, supplied from outset. A finite temperature phase diagram is qualitatively similar to the ones shown in (a) for a fixed value of $t$, since the transition temperature is proportional to the amplitude of the corresponding order-parameter. The temperature scale on the right vertical axis in (a) is arbitrary. The white region at finite-$t$ represents the normal state (a correlated Dirac metal). The coexisting regime in panel (a) arises due to an SO(5) symmetry among two order-parameters~\cite{SO5_comment}. To construct the phase diagram for even larger $\mu$ one should account for higher gradient terms, besides the Dirac components~\cite{comment-highergradient}. 
}~\label{phasediagram}
\end{figure}

This zoo can further be diversified by introducing a relative twist between two layers of graphene, which generically gives rise to incommensurate lattices. In particular, a small twist in bilayer graphene yields nearly flat bands (NFBs) at a series of so-called ``magic angles", first of which occurs at $\theta\sim1.05^\circ$ and \emph{slow} masslesss Dirac fermions residing near the charge-neutrality point (CNP) provide an excellent starting point to describe this system~\cite{castro-neto-PRL2007, morell, macdonald, santos, magaud, roy-yang, kaxiras,senthil}. Such NFBs, reported in recent experiments~\cite{CaoPRL, Cao-Nature1}, represent an ideal arena for the interaction effects to set in~\cite{koponin, roy-single, us,esquinazi,iglovikov,gonzalez-arraga}. In fact, superconductivity with a critical temperature $T_c \sim 1.7$ K has been observed in twisted bilayer graphene close to the first magic angle (MA-TBLG) upon doping this system~\cite{Cao-Nature2}, standing as the first example of pure carbon-based two-dimensional superconductor. Even more intriguingly, the superconductivity possibly arises from a parent Mottlike insulating state~\cite{Cao-Nature2}. Motivated by these experimental observations, we theoretically address the effects of strong electronic interactions on \emph{slow} (due to small Fermi velocity) Dirac fermions~\cite{comment_eeinteractionSC}, constituting an effective low-energy model for MA-TBLG~\cite{castro-neto-PRL2007, morell, macdonald, santos, magaud, roy-yang, kaxiras,senthil}, and arrive at the representative phase diagrams, shown in Figs.~\ref{RG:phaseDiagram} and ~\ref{phasediagram}.

Our main findings can be summarized as follows. First, assuming that the onsite Hubbard repulsion is the dominant finite range Coulomb interaction at the lattice scale, we show that the leading instability of massless slow Dirac fermions near the CNP (zero doping) is toward the formation of an antiferromagnetic (AFM) ground state. However, as the system is doped away from the CNP, pairing interactions develop from the Hubbard repulsion. Among various possible superconducting channels, we show that a time-reversal symmetry breaking valley pseudospin triplet (but spin-singlet) topological $p+ip$ paired state is energetically most favored. Our proposed phase diagrams display such competition (Fig.~\ref{RG:phaseDiagram}) and a possible coexistence of the AFM and paired states (Fig.~\ref{phasediagram}). These features are in qualitative agreement with the recent experimental findings~\cite{Cao-Nature2}. The predicted time-reversal symmetry breaking $p+ip$ pairing can support \emph{quantized} anomalous spin and thermal Hall conductivities, as well as finite polar Kerr and Faraday rotations.

\emph{Model:} Low-energy excitations in a MA-TBLG near the CNP can be described as a collection of slow Dirac fermions~\cite{castro-neto-PRL2007, morell, macdonald, santos, magaud, roy-yang, kaxiras,senthil}, with Fermi velocity $v_{_F}$ being $\sim 25$ times smaller than that in an isolated monolayer graphene~\cite{Cao-Nature1}. The corresponding Dirac Hamiltonian reads as~\cite{Comment-triangular}  
\begin{equation}\label{eq:DiracTBLG}
H_D({\bf k})=v_{_F} \; \sigma_0 \left( \Gamma_{01} k_x + \Gamma_{02} k_y \right)-\mu.
\end{equation}
The spinor basis is $\Psi^\top({\bf k})=\left[ \Psi^\top_\uparrow({\bf k}), \Psi^\top_\downarrow({\bf k}) \right]$, with $\Psi^\top_\sigma({\bf k})=[a_{1,\sigma}({\bf k}),b_{1,\sigma}({\bf k}),a_{2,\sigma}({\bf k}),b_{2,\sigma}({\bf k})]$, where $a,b$ are two sublattices, $1,2$ represent two inequivalent valleys of the single layer graphene lattice, and $\sigma=\uparrow, \downarrow$ are two projections of electronic spin. The chemical potential ($\mu$) is measured from the CNP. Two valleys are located at the corners of the hexagonal Brillouin zone (the ${\bf K}$ points). An identical Hamiltonian describes the low-energy excitations in the other layer and we treat layer as a \emph{decoupled} flavor degree of freedom. The above Hamiltonian arises in the small angle approximation where at low energies \emph{intra-valley} tunneling processes between the layers dominate~\cite{castro-neto-PRL2007,macdonald}. The five mutually anti-commuting four-component $\Gamma$-matrices are the following: $\Gamma_0=\alpha_0 \beta_3$, $\Gamma_1=\alpha_3 \beta_2$, $\Gamma_2=\alpha_0\beta_1$, $\Gamma_3=\alpha_1 \beta_2$ and $\Gamma_5=\alpha_2 \beta_2$. In addition, ten commutators are defined as $\Gamma_{jk}=\left[ \Gamma_j, \Gamma_k \right]/(2i)$. Two sets of Pauli matrices $\{\alpha_\mu\}$ and $\{\beta_{\mu}\}$ respectively operate on the valley or \emph{pseudo-spin} and sublattice degrees of freedom, and $\left\{ \sigma_\mu \right\}$ on the spin indices. The noninteracting theory enjoys separate $SU(2)$ pseudo-spin and spin rotational symmetries, respectively generated by $\left\{ \Gamma_3, \Gamma_5, \Gamma_{35} \right\}$ and $\left\{ \sigma_1, \sigma_2, \sigma_3 \right\}$, besides the rotational and translational symmetries, respectively generated by $\Gamma_{12}$ and $\Gamma_{35}$, see also Supplementary Materials (SM)~\cite{supplementary}.

The quintessential features of an interacting Dirac liquid in MA-TBLG can be captured by the Hubbard model
\begin{equation}
H_U =\frac{U}{2} \sum_{i} n_{i,\uparrow} \; n_{i,\downarrow},
\end{equation}
as the onsite interaction is the dominant component of finite range Coulomb interaction~\cite{katsnelson}, where $U (>0)$ denotes the strength of the onsite repulsion, and $n_{i,\sigma}$ is the number operator at site $i$ with spin-projection $\sigma=\uparrow, \downarrow$. The low-energy version of the Hubbard model can be obtained by decomposing the fermionic operators in terms of the Fourier modes around two valleys, leading to
\begin{eqnarray}
H_{U} &=& \int d^2 {\bf x} \bigg[ g^t_{_1} \left( \Psi^\dagger {\boldsymbol \sigma} \; {\mathbb 1} \Psi \right)^2 + g^t_{_2} \left( \Psi^\dagger {\boldsymbol \sigma} \; \Gamma_0 \Psi \right)^2 \nonumber \\
&+& g^t_{_3} \sum_{j=1,2} \:\: \sum_{{k=3,5}} \left( \Psi^\dagger {\boldsymbol \sigma} \; \Gamma_{jk} \Psi \right)^2 \bigg],
\end{eqnarray}
where $g^t_1=g^t_2=2g^t_3=-U/12$, see SM~\cite{supplementary}. While the first two terms represent \emph{forward} scattering, the last one corresponds to \emph{back-scattering} between two valleys. Using Fierz identity~\cite{roy-goswami-juricic}, the above Hamiltonian can also be written in terms of four-fermion interactions in the spin-singlet channels, after taking $g^t_j \to -3 g^s_j$ and ${\boldsymbol \sigma} \to \sigma_0$. This change of representation, along with the change of signs for all quartic couplings, confirms that repulsive Hubbard interaction is conducive for excitonic orderings, but only in the \emph{spin-triplet} channel.

\emph{Results:} To compare the propensity toward the formation of various spin-triplet excitonic orderings, characterized by an order-parameter $\langle \Psi^\dagger {\boldsymbol \sigma} M \Psi \rangle$, where $M$ is a 4-dimensional Hermitian matrix, we compute corresponding \emph{bare} susceptibility by performing Hubbard-Stratonovich decomposition of all quartic terms appearing in the Hubbard model, and subsequently integrating out fermions. For zero external momentum and frequency the bare susceptibility at $T=0$ reads
\begin{align}
\chi= -\int \frac{d\omega \; d^2 {\bf k}}{(2 \pi)^3} \: {\bf Tr} \left[ G_{\bf k}(i \omega) \sigma_j M G_{\bf k}(i \omega) \sigma_j M \right],  
\end{align}
for a specific choice of spin-axis $j=x,y,z$, where $\omega$ is the fermionic Matsubara frequency and $G_{\bf k}(i \omega)$ is the fermionic Green's function. Integral over momentum is restricted up to an ultraviolet cutoff $\Lambda$~\cite{comment_UVcutoff}. The bare susceptibility is \emph{largest} for $M=\Gamma_0$, as this matrix operator anti-commutes with the Dirac Hamiltonian (for $\mu=0$), and $\langle \Psi^\dagger {\boldsymbol \sigma} \Gamma_0 \Psi \rangle$ represents the AFM order in honeycomb lattice. Therefore, a mean-field analysis of the Hubbard model indicates an onset of AFM order in a half-filled MA-TBLG, which is quite natural as the honeycomb lattice does not offer any frustration for a staggered arrangement of electronic spin~\cite{herbut, sorella, assaad, foster}. Even though suppression of Fermi velocity in MA-TBLG near the CNP boosts the propensity toward the formation of anti-ferromagnet, since $\chi \sim \Lambda/v_{_F} \sim E_\Lambda/v^2_{_F}$, where $E_\Lambda \approx v_{_F} \Lambda$ is the Dirac band width~\cite{comment_UVcutoff}, due to the vanishing density of states (DoS) such ordering always takes place at a finite coupling. Presently it is not clear whether the Hubbard-U is sufficient to nucleate the AFM order near the CNP (due to reduced $E_\Lambda$). But, recent experiment is suggestive of a metallic phase in its vicinity~\cite{Cao-Nature1, Cao-Nature2, comment-highergradient}. While the metallic phase can be an AFM, for large enough $U$ it can become a Mott insulator~\cite{sorella, assaad}.

Yet another (perhaps most exciting) experimental observation in Ref.~\cite{Cao-Nature2}, is the onset of superconductivity with increasing carrier density in the system. Next we seek to understand a possible microscopic origin of such a paired state. To facilitate the following discussion, we now introduce a 16-component Nambu-doubled basis as $\Psi^\top_N =\left[\Psi^\top_{\sigma}, i \sigma_2 \Gamma_{15} \left(\Psi^{\ast}_{\sigma}\right)^\top \right]$. Generalizing the Fierz identity for all possible four-fermion pairing interactions~\cite{vafek}, given by $\sum_{j} g^p_j \left( \Psi^\dagger_N \eta_1 M_j \Psi_N \right)^2$, where the summation over $j$ runs over all superconducting vertices and $M_j$s are eight-dimensional Hermitian matrices, we find that onsite repulsive Hubbard interaction does not favor any superconductivity, at least when $\mu=0$, since then $g^p_j \geq 0$ for all $j$. The newly introduced Pauli matrices $\left\{ \eta_\mu \right\}$ operate in the Nambu (particle-hole) space. However, at finite chemical doping, a pairing interaction, conducive for superconductivity, may arise from \emph{incipient} AFM fluctuations (without any long-range order)~\cite{AFM_fluctuation}.

The mechanism of incipient fluctuation mediated pairing interaction can be demonstrated from the \emph{vertex} correction ($\delta g^p_{j,{\rm v}}$) to the pairing interaction $g^p_j$ due to a dominant underlying AFM interaction ($g^t_{_2}$), given by
\allowdisplaybreaks[4]
\begin{eqnarray}
\delta g^p_{j,{\rm v}} &=& g^t_{_2} \left( \Psi^\dagger_N \eta_1 M_j \Psi_n \right)  \Psi^\dagger_N \bigg[\sum_{i \omega_n} \int_{{\bf k}} {\boldsymbol \sigma} \Gamma_0 \:  G^{N}_{{\bf k}}(i\omega_n, \mu) \nonumber \\
&\times& \eta_1 M_j \: G^{N}_{{\bf k}}(i\omega_n, \mu) \: {\boldsymbol \sigma} \Gamma_0 \bigg] \Psi_N,          		
\end{eqnarray}		
where $G^{N}_{{\bf k}}(i\omega, \mu)$ is the fermionic Green's function in the Nambu-doubled basis at finite chemical potential ($\mu$). Notice that $\delta g^p_{j,{\rm v}}<0$ (thus conducive for superconductivity) only when $\left\{ M_j, {\boldsymbol \sigma} \Gamma_0 \right\}=0$, since for repulsive Hubbard interaction $g^t_{_2}<0$. Such criterion immediately justifies one feature: \emph{Incipient magnetic fluctuations can only favor spin-singlet pairing}. Among many possible spin-singlet pairings only \emph{two}, belonging to two-dimensional $E_g$ and $A_{1{\bf k}}$ representations of $D_{3d}$ point group (since we trade the influence of emergent triangular lattice in favor of a renormalized Dirac liquid), satisfy the above stringent criterion. In the announced Nambu-doubled basis these two pairings respectively take the following form: (1) $E_g$: $\left( \eta_1, \eta_2 \right) \sigma_0 \left( \Gamma_1, \Gamma_2 \right)$, (2) $A_{1{\bf k}}$: $\left( \eta_1, \eta_2 \right) \sigma_0 \left( \Gamma_{03}, \Gamma_{05} \right)$. The $E_g$ pairing directly couples two inequivalent valleys and preserves the translational symmetry in the paired state. This paired state however breaks the rotational symmetry, since $\{\Gamma_j, \Gamma_{12} \}=0$ for $j=1,2$, and stands as an example of \emph{nematic} superconductor. By contrast, the $A_{1{\bf k}}$ pairing takes place separately near two valleys at $\pm {\bf K}$ and breaks the translational symmetry, since $\{\Gamma_{0j}, \Gamma_{35} \}=0$ for $j=3,5$, with periodicity of $2|{\bf K}|$. It represents a \emph{commensurate} Fulde-Farrel-Larkin-Ovchinikov (FFLO) pairing~\cite{roy-solo}, also known as \emph{pair-density-wave}. Enhancement of pairing interactions in these two channels stems from an underlying SO(5) symmetry among AFM and superconducting orders~\cite{SO5_comment}.

The $A_{1{\bf k}}$ pairing commutes with the Dirac Hamiltonian, implying that $\delta g^p_{A_{1{\bf k}},{\rm v}}=0$ when $\mu=0$. By contrast, $\delta g^p_{E_g,{\rm v}}<0$ for repulsive Hubbard interaction. Hence, the dominant propensity in close proximity to an AFM ordering is in the $E_g$ channel. This conclusion remains valid even when $\mu \neq 0$. The competition between the anti-ferromagnet and $E_g$ pairing at finite $T$ and $\mu$ can be appreciated from the following leading order renormalization group (RG) flow equations for the dominant coupling of the Hubbard model ($g^t_{_2}$) and the source terms for these two channels (respectively denoted by $m$ and $\Delta$)~\cite{interband_comment}
\begin{align}~\label{RG:finiteT}
&\frac{dg^t_{_2}}{d \ell} = -g^t_{_2} + 4 \left( g^t_{_2} \right)^2 f(t,\mu), \:\: \frac{d t}{d\ell}=t, \: \frac{d \mu}{d\ell}=\mu, \nonumber \\  
&\frac{d \ln m}{d\ell}-1 = \frac{7g^t_{_2}}{2} f(t,\mu), \:\:
\frac{d \ln \Delta}{d\ell}-1= \frac{3g^t_{_2}}{4} h(t,\mu),
\end{align} 
obtained by performing a summation over fermionic Matsubara frequencies and subsequently integrating out a thin Wilsonian shell $\Lambda e^{-\ell} < |{\bf k}| < \Lambda$. Two lengthy functions $f(t,\mu)$ and $h(t,\mu)$ are shown in SM~\cite{supplementary}. RG flow equations are expressed in terms of dimensionless variables obtained by taking $g^t_{_2} \Lambda/(\pi v_{_F}) \to g^t_{_2}$, $(T,\mu,m,\Delta)/(\Lambda v_{_F}) \to (t,\mu,m,\Delta)$. To the leading order $v_{_F}$ does not get renormalized by local interactions.

Both temperature and chemical potential introduce infra-red cutoffs for the RG flow, respectively given by $\ell^t_\ast=-\ln t(0)$ and $\ell^\mu_\ast=- \ln \mu(0)$, where $X(0)(<1)$ denotes bare dimensionless parameters. Therefore, we run the flow of $g^t_2$ only up to a scale $\ell_\ast=\min.(\ell^t_\ast, \ell^\mu_\ast)$. Now depending on the bare coupling constant $g^t_{_2}(0)$ two scenarios can arise: (a) $g^t_{_2}(\ell_\ast)<1$ or (b) $g^t_{_2}(\ell_\ast)>1$. While the former situation describes an interacting Dirac liquid without any spontaneous symmetry breaking, the latter one indicates breakdown of perturbation theory and onset of an ordered phase. To determine the actual pattern of the symmetry breaking we simultaneously run the flow of the source terms for two competing channels. When $g^t_2(\ell_\ast)>1$ the broken symmetry phase represents anti-ferromagnet if $m(\ell_\ast)> \Delta(\ell_\ast)$ or an $E_g$ superconductor if $\Delta(\ell_\ast)> m(\ell_\ast)$. Following this procedure, we construct various cuts of the phase diagram in the $\left(g^t_2,t \right)$ plane, for different values of $\mu$, shown in Fig.~\ref{RG:phaseDiagram}.

The phase diagram shows that beyond a critical strength of interaction and at zero doping the only possible ordered phase is anti-ferromagnet. But, with increasing chemical doping the critical interaction strength for the AFM ordering increases, while a superconducting phase develops for a weaker interaction. Presently it is unknown how the bare strength of the Hubbard interaction scales as a function of the chemical doping, which can only be accomplished from microscopic calculation~\cite{katsnelson}. Nevertheless, it is conceivable for the real system to follow a path in the phase diagram (Fig.~\ref{RG:phaseDiagram}) that goes through the superconducting phase for low chemical doping, entering into the AFM phase at higher doping, as we show (phenomenologically) below (Fig.~\ref{phasediagram}).

To demonstrate a possible coexistence of anti-ferromagnet and $E_g$ pairing inside the ordered phase, where the perturbative RG breaks down (since $g^t_{_2}>1$), we focus on the following mean-field free-energy~\cite{comment_phaselocking} 
\begin{equation}~\label{free-energy:Eq}
\bar{f}= \frac{m^2}{2 g_{_1}} + \frac{|\Delta|^2}{4 g_{_2}}- 4 t \sum_{j=1,2} \int^{\prime} \frac{d^2{\bf k}}{(2 \pi)^2} \ln \left[2 \cosh \left( \frac{E_j}{2T} \right) \right],
\end{equation}
for $k_B=1$. All quantities in $\bar{f}$ are dimensionless, with 
\begin{eqnarray}
 E_j &=& \big\{ v^2_F k^2 + \mu^2 + m^2 +|\Delta|^2 + (-1)^{j} \big[ 2 v^2_F k^2 |\Delta|^2  \nonumber \\
&\times& (1-\sin 2\theta_{\bf k}) + 4 (v^2_F k^2 + m^2) \mu^2 \big]^{1/2}  \big\}^{1/2},
\end{eqnarray}
where $\theta_{\bf k}=\tan^{-1}(k_y/k_x)$. The coupling constant $g_{_1}$($g_{_2}$) supports AFM (superconducting) order. Minimization of $\bar{f}$ with respect to $m$ and $\Delta$ leads to coupled gap equations, shown in the SM~\cite{supplementary}, which we solve numerically to arrive at the phase diagram, displayed in Fig.~\ref{phasediagram}. We find that it is possible for the system to first enter into a superconducting phase as we increase the chemical doping away from the CNP and subsequently support an AFM state, in qualitative agreement with experimental observations~\cite{Cao-Nature2}. The underlying SO(5)symmetry between these two orders also permits a region of coexistence.

\emph{Responses $\&$ topology:} To appreciate the phase locking between two components of the $E_g$ pairing and emergent topology of BdG quasi-particles deep inside the paired state, we now project it onto the Fermi surface, yielding the following effective single-particle Hamiltonian 
\begin{equation}
H_{\rm BdG}= \xi_{\bf k} \eta_3 + \frac{|\Delta|}{k_F} \bigg[ \eta_1 \sigma_0 k_x + \eta_2 \sigma_0 k_y \bigg] \alpha_3
\end{equation}
where $\xi_{\bf k}= v  k-\mu$ or $\left[ k^2/(2 m)-\mu_\ast \right]$ respectively in the presence or absence of AFM order, $k_F$ is Fermi momentum and $\mu_\ast=\mu-m$, see SM~\cite{supplementary}. The phase locking between the $k_x$ and $k_y$ components is dictated by the requirement of a \emph{maximally gapped} Fermi surface (since all entries in $H_{\rm BdG}$ mutually anticommute), yielding largest gain of condensation energy. The time-reversal symmetry is spontaneously broken in this paired state. Hence, the $E_g$ pairing close to the Fermi surface assumes the form of a \emph{fully gapped} topological $p+ip$ pairing. This is a class D \emph{spin-singlet}, but \emph{pseudo- or valley-spin triplet} pairing, characterized by Z topological invariant~\cite{tenfold}, and supports \emph{quantized} anomalous spin and thermal Hall conductivities, respectively given by
\begin{eqnarray}
\sigma^0_{xy, S}= \frac{\hbar}{4 \pi}, \quad
\lim_{T \to 0} \frac{\kappa_{xy}}{T} = \frac{2}{3} \frac{\pi^2 k^2_B}{h},
\end{eqnarray}
as temperature $T \to 0$, respectively, as well as finite polar Kerr and Faraday rotations~\cite{kapitulnik}. By contrast, if the pairing interaction exists over the entire valence and conduction bands, then a maximally gapped Fermi surface, comprised of two point nodes around which DoS vanishes as $\varrho(E) \sim |E|$, is obtained from a \emph{time-reversal symmetric} combination $|\Delta| \eta_1 \sigma_0 \left( \Gamma_1 \cos{\phi} + \Gamma_2 \sin \phi \right)$ of the $E_g$ pairing. Here, $\phi$ is an internal angle, which should be locked at specific values, depending on the underlying crystal potential. Therefore, above mentioned responses (spin and thermal Hall conductivities, Kerr and Faraday rotations) can probe the nature or extent of the pairing interaction in MA-TBLG. To this end measurement of the penetration depth ($\Delta \lambda$) can be instrumental leading to $\Delta \lambda \sim (T/T_c)^{n}$, with $n \approx 3-4$ for fully gapped $p+ip$ state, but $n=1$ for gapless $E_g$ pairing, when $T \ll T_c$~\cite{prozorov-review}.

In brief, starting from an effective low-energy model of interacting \emph{slow} Dirac fermions, we demonstrate that MA-TBLG can display an intriguing confluence of competing AFM and singlet $E_g$ nematic superconducting phases~\cite{singlet_comment}, if the onsite Hubbard repulsion stands as the dominant component of finite range Coulomb interaction. Close to the Fermi surface $E_g$ superconductor represents a pseudo-spin triplet $p+ip$ pairing, which can support quantized anomalous spin and thermal Hall conductivities. Such a superconducting state supports Majorana edge modes~\cite{volovik-book, read-green}, turning MA-TBLG into a potential platform for topological quantum computing~\cite{dassarma-RMP}.

\emph{Acknowledgments}: We are thankful to Pablo Jarillo-Herrero and Andras Szab$\acute{\mbox{o}}$ for useful discussions.



\begin{thebibliography}{}

\bibitem{CNT-book}  A. Jorio, G. Dresselhaus, M. S. Dresselhaus, \emph{Carbon Nanotubes: Advanced Topics in the Synthesis, Structure, Properties and Applications} (Spinger, Berlin, 2008).

\bibitem{katsnelson-book} M. I. Katsnelson, \emph{Graphene: Carbon in Two Dimensions} (Cambridge University Press, Cambridge, U. K., 2012)

\bibitem{Grigorieva-Geim-Nature2013} A. K. Geim and I. V. Grigorieva, Nature {\bf 499}, 419 (2013).

\bibitem{graphene-discovery} K. S. Novoselov, A. K. Geim, S. V. Morozov, D. Jiang, Y. Zhang, S. V. Dubonos, I. V. Grigorieva, and A. A. Firsov, Science {\bf 306}, 666 (2004).

\bibitem{ballistic-graphene} K. S. Novoselov, A. K. Geim, S. V. Morozov, D. Jiang, M. I. Katsnelson, I. V. Grigorieva, S. V. Dubonos, and A. A. Firsov, Nature (London) {\bf 438}, 197 (2005).

\bibitem{Kim-qhe-monolayer} Y. Zhang, Y. Tan, H. L. Stormer, and P. Kim, Nature (London) {\bf 438}, 201 (2005).

\bibitem{bilayer-qhe} K. S. Novoselov, E. McCann, S. V. Morozov, V. I. Falko, M. I. Katsnelson, U. Zeitler, D. Jiang, F. Schedin, and A. K. Geim, Nat. Phys. {\bf 2}, 177 (2006).

\bibitem{zhang2011} L. Zhang, Y. Zhang, J. Camacho, M. Khodas, and Igor Zaliznyak, Nat. Phys. {\bf 7}, 953 (2011).

\bibitem{castro-neto-PRL2007} J. M. B. Lopes dos Santos, N. M. R. Peres, and A. H. Castro Neto, Phys. Rev. Lett. {\bf 99}, 256802 (2007).

\bibitem{morell} E. Suarez Morell, J. D. Correa, P. Vargas, M. Pacheco, and Z. Barticevic, Phys. Rev. B {\bf 82}, 121407 (2010).

\bibitem{macdonald}  R. Bistritzer and A. H. MacDonald, Proc. Natl. Acad. Sci. U. S. A.  {\bf 108}, 12233 (2011).

\bibitem{santos} J. M. B. L. dos Santos, N. M. R. Peres, and A. H. Castro Neto,  Phys. Rev. B {\bf 86}, 155449 (2012).

\bibitem{magaud} G. T. de Laissardiére, D. Mayou, and L. Magaud, Phys. Rev. B {\bf 86} 125413 (2012).

\bibitem{roy-yang} B. Roy and K. Yang, Phys. Rev. B {\bf 88}, 241107(R) (2013).

\bibitem{kaxiras} S. Fang and E. Kaxiras, Phys. Rev. B {\bf 93}, 235153 (2016).

\bibitem{senthil} H-C. Po, L. Zou, A. Vishwanath, T. Senthil, Phys. Rev. X {\bf 8}, 031089 (2018). 

\bibitem{CaoPRL} Y. Cao, J. Y. Luo, V. Fatemi, S. Fang, J. D. Sanchez-Yamagishi, K. Watanabe, T. Taniguchi, E. Kaxiras, and P. Jarillo-Herrero, Phys. Rev. Lett. {\bf 117}, 116804 (2016).

\bibitem{Cao-Nature1} Y. Cao, V. Fatemi, A. Demir, S. Fang, S. L. Tomarken, J. Y. Luo, J. D. Sanchez-Yamagishi, K. Watanabe,
T. Taniguchi, E. Kaxiras, R. C. Ashoori, and P. Jarillo-Herrero, Nature (London) {\bf 556}, 80 (2018).

\bibitem{koponin} N. B. Kopnin, T. T. Heikkil\"a, and G. E. Volovik, Phys. Rev. B {\bf 83}, 220503(R) (2011).

\bibitem{us} B. Roy and V. Juri\v ci\' c, Phys. Rev. B {\bf 90}, 041413(R) (2014).

\bibitem{esquinazi} P. Esquinazi, T. T. Heikkila, Y. V. Lysogorskiy, D. A. Tayurskii, and G. E. Volovik, JETP Lett. {\bf 100}, 336 (2014).

\bibitem{iglovikov} V. I. Iglovikov, F. Hebert, B. Gremaud, G. G. Batrouni, and R. T. Scalettar, Phys. Rev. B {\bf 90}, 094506 (2014).

\bibitem{roy-single} B. Roy, Phys. Rev. B {\bf 96}, 041113(R) (2017). 

\bibitem{gonzalez-arraga} L. A. Gonzalez-Arraga, J. L. Lado, F. Guinea, and P. San-Jose, Phys. Rev. Lett. {\bf 119}, 107201 (2017).

\bibitem{Cao-Nature2} Y. Cao, V. Fatemi, S. Fang, K. Watanabe, T. Taniguchi, E. Kaxiras, and P. Jarillo-Herrero, Nature {\bf 556}, 43 (2018).

\bibitem{comment_eeinteractionSC} $T_c/T_F \approx 0.06$ in MA-TBLG (see Fig.~6 of Ref.~\cite{Cao-Nature2}), where $T_F$ is the Fermi temperature, quite comparable to other strongly correlated superconducting materials, such as cuprates, pnictides and heavy-fermion compounds, suggesting that superconductivity in MA-TBLG is possibly driven by electronic interactions~\cite{pasupathy}.  


\bibitem{pasupathy} A. Kerelsky, L. McGilly, D. M. Kennes, L. Xian, M. Yankowitz, S. Chen, K. Watanabe, T. Taniguchi, J. Hone, C. Dean, A. Rubio, A. N. Pasupathy, arXiv:1812.08776 

\bibitem{supplementary} See Supplemental Materials at XXX-XXXX for the details of (a) Renormalization groups calculation, (b) competition between AFM and $E_g$ pairing in mean-field limit, (c) band diagonalization in the presence of an underlying pairing. 

\bibitem{Comment-triangular} Since effective lattice constant of an emergent triangular lattice, formed by the localized AA regions, is $a_{\rm SL} \sim 200 \mathring{\rm A}$ (Moir\'e superlattice constant), is much bigger than typical superconducting and magnetic coherence lengths $\sim 10 \mathring{\rm A}$, such localized states can only renormalize the various parameters (such as Fermi velocity, Hubbard interaction) of itinerant fermions in MA-TBLG.  


\bibitem{katsnelson} T. O. Wehling, E. \ifmmode \mbox{\c{S}}\else \c{S}\fi{}a\ifmmode \mbox{\c{s}}\else \c{s}\fi{}\ifmmode \imath \else \i \fi{}o\ifmmode \breve{g}\else \u{g}\fi{}lu, C. Friedrich, A. I. Lichtenstein, M. I. Katsnelson, and S. Bl\"ugel, Phys. Rev. Lett. {\bf 106}, 236805 (2011).

\bibitem{roy-goswami-juricic}  See for example, B. Roy, P. Goswami, V. Juri\v ci\' c, Phys. Rev. B {\bf 95}, 201102(R) (2017).

\bibitem{comment_UVcutoff} In MA-TBLG the ultraviolet cutoff $\Lambda \sim a^{-1}_{\rm SL}$~\cite{Comment-triangular}.  

\bibitem{herbut} I. F. Herbut, Phys. Rev. Lett. {\bf 97} 146401 (2006).

\bibitem{sorella} S. Sorella, Y. Otsuka, and S. Yunoki, Sci. Rep. {\bf 2}, 992 (2012).

\bibitem{assaad} F. F. Assaad and I. F. Herbut, Phys. Rev. X {\bf 3}, 031010 (2013).

\bibitem{foster} B. Roy and M. S. Foster, Phys. Rev. X {\bf 8}, 011049 (2018).

\bibitem{comment-highergradient} Sufficiently far from the CNP, higher gradient terms $\sim |{\bf k}|^2, |{\bf k}|^3, \cdots$, become important, which, due to higher DoS can suppress the irrelevance of interaction (arising from vanishing DoS of pure Dirac fermions) and enhance the propensity toward various orderings. A systematic inclusion of higher gradient terms, which does not qualitatively alter our results, is left for a future investigation.

\bibitem{vafek} J. M. Murray and O. Vafek, Phys. Rev. B {\bf 89}, 205119 (2014).

\bibitem{AFM_fluctuation} This mechanism of generating pairing interaction is distinct from conventional magnon-driven superconductivity in the presence of long-range AFM order. Here superconductivity is driven by incipient or short-range AFM ordering, lacking long-range phase locking.  


\bibitem{roy-solo} B. Roy, Phys. Rev. B {\bf 88}, 075415 (2013).

\bibitem{SO5_comment} Explicitly, SO(5) order-parameter vector is constituted by $\left\{ \sigma_1 \Gamma_0, \sigma_2 \Gamma_0, \sigma_3 \Gamma_0, \Gamma_j, \Gamma_k \right\}$, where either (a) $j=1,k=2$ (for $E_g$ pairing), or (b) $j=03,k=05$ (for $A_{1 {\bf k}}$ pairing).

\bibitem{interband_comment} Since in MA-TBLG $U \sim$ band width, both inter-band and intra-band scatterings are equally important, even when chemical potential is placed away from the CNP.

\bibitem{comment_phaselocking} We choose a specific realization (with $\theta=0=\phi$) of $E_g$ pairing from its most general configuration: $(\eta_1 \cos \theta + \eta_2 \sin \theta) \sigma_0 (\Gamma_1 \cos \phi + \Gamma_2 \sin \phi)$. Due to a $U(1) \otimes U(1)$ symmetry of the order-parameter any choice of $\theta$ and $\phi$ will produce a phase diagram, similar to the one in Fig.~\ref{phasediagram}. 

\bibitem{tenfold} A. P. Schnyder, S. Ryu, A. Furusaki, A. W. W. Ludwig, Phys. Rev. B {\bf 78}, 195125 (2008).

\bibitem{kapitulnik} A. Kapitulnik, J. Xia, E. Schemm, and A. Palevski, New J. Phys. {\bf 11}, 055060 (2009).

\bibitem{prozorov-review} R. Prozorov, R. W. Giannetta, Supercond. Sci. Technol. {\bf 19}, R41 (2006).

\bibitem{singlet_comment} The upper critical field for superconducting in MA-TBLG is $B^c_\perp \approx 0.4 T \to 0.4$K (with $g$-factor $\approx 2$ for electrons in graphene), much smaller than $T_c \approx 1.7$K~\cite{Cao-Nature2}, suggesting the paired state is possibly a spin-singlet.  

\bibitem{volovik-book} G. E. Volovik, \emph{The Universe in a Helium Droplet} (Clarendon, Oxford, U. K., 2003).

\bibitem{read-green} N. Read and D. Green, Phys. Rev. B {\bf 61}, 10267 (2000).

\bibitem{dassarma-RMP} C. Nayak, S. H. Simon, A. Stern, M. Freedman, and S. Das Sarma, Rev. Mod. Phys. {\bf 80}, 1083 (2008).


\end{thebibliography}
\end{document}